\documentclass[11pt,letterpaper]{article}

\usepackage{mathrsfs}           
\usepackage{vmargin,fancyhdr}   
\usepackage{algorithm}
\usepackage{algorithmic}

\usepackage{enumerate}

\usepackage{amsmath,amssymb}    
\usepackage{verbatim}           
\usepackage{xspace}             
\usepackage{graphicx,float}     
\usepackage{ifthen,calc}        
\usepackage{textcomp}           
\usepackage{fancybox}           
\usepackage{hhline}             
\usepackage{float}              

\usepackage[rflt]{floatflt}     
\usepackage[small,compact]{titlesec}
\usepackage{setspace}
\usepackage{subfigure}

\usepackage{color}
\definecolor{ForestGreen}{rgb}{0.1333,0.5451,0.1333}
\usepackage[letterpaper,dvips,
            colorlinks,linkcolor=ForestGreen,citecolor=ForestGreen,
            backref,
            bookmarks,bookmarksopen,bookmarksnumbered]
           {hyperref}
\usepackage[dvips]{thumbpdf}

\setpapersize{USletter}
\setmarginsrb{.75in}{.5in}        
             {.75in}{.5in}        
             {.25in}{.25in}     
             {.25in}{.5in}      
\newcommand{\showccc}[0]{0}
\newcommand{\ccc}[2][nothing]{
  \ifthenelse{\showccc=0}{}{
    \ensuremath{^{\Lsh\Rsh}}\marginpar{\raggedright\tiny\textsf{%
        \ifthenelse{\equal{#1}{nothing}}{}{\textbf{#1}\\}#2}}}}

\newcounter{hours}\newcounter{minutes}
\newcommand{\hhmm}{%
  \setcounter{hours}{\time/60}%
  \setcounter{minutes}{\time-\value{hours}*60}%
  \ifthenelse{\value{hours}<10}{0}{}\thehours:%
  \ifthenelse{\value{minutes}<10}{0}{}\theminutes}
\ifthenelse{\showccc=0}{\rhead{}}{\rhead{\today \ [\hhmm]}}
\newtheorem{theorem}{Theorem}[section]
\newtheorem{claim}[theorem]{Claim}

\newtheorem{corollary}[theorem]{Corollary}
\newtheorem{definition}[theorem]{Definition}

\newtheorem{lemma}[theorem]{Lemma}

\newcommand{\Proof}[0]{\smallskip\noindent\textit{\textbf{Proof}}\quad}

\newcommand{\QED}[0]{\hfill\ensuremath{\blacksquare}\medspace}

\newcommand{\R}[0]{\ensuremath{\mathbb{R}}}

\floatstyle{ruled}
\newfloat{algo}{tbp}{lop}
\floatname{algo}{Algorithm}

\usepackage{float}
\floatstyle{plain}\newfloat{myfig}{t}{figs}[section]
\floatname{myfig}{\textsc{Figure}}
\floatstyle{plain}\newfloat{myalg}{H}{algs}[section]
\floatname{myalg}{}
\begin{document}

\title{A nearly-$m \log n$ time solver for SDD linear systems \thanks{Partially supported by the National Science Foundation under grant number CCF-1018463.}}
\author{Ioannis Koutis \\
CSD-UPRRP \\ ioannis.koutis@upr.edu   \and Gary L. Miller \\  CSD-CMU  \\ glmiller@cs.cmu.edu \and Richard Peng \\  CSD-CMU \\yangp@cs.cmu.edu}
\maketitle

\begin{abstract}
We present an improved algorithm for solving symmetrically diagonally
dominant linear systems.
On input of an $n\times n$ symmetric diagonally dominant matrix $A$ with $m$ non-zero entries and a vector $b$ such that $A\bar{x} = b$ for some (unknown) vector $\bar{x}$,  our algorithm computes a vector $x$ such that $||{x}-\bar{x}||_A < \epsilon ||\bar{x}||_A $ \footnote{$||\cdot||_A$ denotes the A-norm} in time $${\tilde O}(m\log n \log (1/\epsilon)).\footnote{The $\tilde{O}$ notation hides a $(\log\log n)^2$ factor}$$

\vspace{0.05cm}

The solver utilizes in a standard way a `preconditioning' chain of progressively sparser graphs. To claim the faster running time we make a two-fold improvement in the algorithm for constructing the chain. The new chain exploits previously unknown properties of the graph sparsification algorithm given in [Koutis,Miller,Peng, FOCS 2010], allowing for stronger preconditioning properties. We also present an algorithm of independent interest that constructs nearly-tight low-stretch spanning trees in time $\tilde{O}(m\log{n})$, a factor of $O(\log{n})$ faster than the algorithm in [Abraham,Bartal,Neiman, FOCS 2008]. This speedup directly reflects on the construction time of the preconditioning chain.

\end{abstract}

\section{Introduction}

Solvers for symmetric diagonally dominant (SDD)\footnote{A system $Ax=b$ is SDD when $A$ is symmetric and $A_{ii} \geq \sum_{j \not= i} |A_{ij}|$.} systems are a crucial component of the fastest known algorithms for a multitude of problems that include
(i) Computing the first non-trivial (Fiedler) eigenvector of the graph,  with well known applications to the sparsest-cut problem \cite{Fiedler73,SpielmanTeng96,chung1};
(ii) Generating spectral sparsifiers that also act as cut-preserving
sparsifiers \cite{SpielmanS08};
(iii) Solving linear systems derived from elliptic finite element discretizations of a significant
class of partial differential equations \cite{BHV04};
(iv) Generalized lossy flow problems \cite{SpielmanDaitch08};
(v) Generating random
spanning trees \cite{KelnerMadry09};
(vi) Faster maximum flow algorithms \cite{FastFlow10};
and
(vii) Several optimization
problems in computer vision \cite{KMST-TR-09,KoutisMT11} and graphics
\cite{mccannreal08,pushkarHarmonic07}.

These algorithmic advances were largely motivated by the seminal work of Spielman and Teng  who gave the first nearly-linear time solver  for SDD systems \cite{SpielmanTeng04,EEST05,SpielmanTeng08c}. The running time of their solver is a large number of polylogarithmic factors away
from the obvious linear time lower bound. In recent work, building upon further work of Spielman and Srivastava \cite{SpielmanS08}, we presented a simpler and faster SDD solver with a run time of $\tilde{O}(m \log^2 n \log \epsilon^{-1})$,
where $m$ is the number of nonzero entries, $n$ is the number of variables, and $\epsilon$ is a standard measure of the approximation error
\cite{KoutisMP10}.

It has been conjectured that the algorithm of \cite{KoutisMP10} is not optimal \cite{Spielman10Survey,Teng10Survey, Spielman10Laplacian}. In this paper we give an affirmative answer by presenting a solver that runs in $\tilde{O}(m \log n \log \epsilon^{-1})$ time.

The $O(\log n)$ speedup of the SDD solver applies to all algorithms listed above, and we believe that it will prove to be quite important in practice, as applications of SDD solvers frequently involve  massive graphs \cite{Teng10Survey}.

\subsection{Overview of our techniques}

The key to all known near-linear work SDD solvers is {\em spectral graph sparsification}, which on a given input graph $G$ constructs a sparser graph $H$ such that $G$ and $H$ are `spectrally similar' in the {\em condition number} sense, defined in Section \ref{sec:definitions}. Spectral graph sparsification can be seen as a significant strengthening of the notion of cut-preserving sparsification \cite{BenczurK96}.

The new solver follows the framework of recursive preconditioned Chebyshev iterations \cite{SpielmanTeng08c,KoutisMP10}. The iterations are driven by a so-called
{\em preconditioning chain} $\{G_1,H_1,G_2,H_2,\ldots,\}$ of graphs, where $H_i$ is a spectral sparsifier for $G_i$ and $G_{i+1}$ is generated by contracting $H_i$ via a greedy elimination of degree 1 and 2 nodes. The total work of the solver includes the time for constructing the chain, and the work spent on actual iterations which is a function on the preconditioning quality of the chain. The preconditioning quality of the chain in turn depends on the guarantees of the sparsification algorithm.

More concretely, all sparsification routines that have been used in SDD solvers conform to the same template; on input a graph $G$ with $n$ vertices and $m$ edges returns a graph $H$ with $n+\tilde{O}(m\log^c n)/\kappa$ edges such that the condition number of the Laplacians of $G$ and $H$ is $\kappa$. In all known SDD solvers the factor $\tilde{O}(\log^c n)$ appears directly in the running time of the SDD solver. In particular the solver of \cite{KoutisMP_FOCS10} was based on a sparsification routine for which $c=2$.

The optimism that SDD systems can be solved in time $\tilde{O}(m \log n \log \epsilon^{-1})$ has mainly been based on the result of  Kolla et al. \cite{Kolla10} who proved that there is a polynomial (but far from nearly-linear) time algorithm that returns a sparsifier with $c=1$. However, our new solver is instead based on a slight modification and a deeper analysis of the sparsification algorithm in \cite{KoutisMP10} which enables a subtler chain construction.

The incremental sparsification algorithm in \cite{KoutisMP10} computes and keeps in $H$ a properly scaled copy of a low-stretch spanning tree of $G$, and adds to $H$ a number of off-tree samples from $G$. The key enabling observation in the new analysis is that the total stretch of the off-tree edges is essentially invariant under sparsification. In other words, the total stretch of the off-tree edges in $H_i$ is at most equal to that $G_i$. The total stretch is invariable under the graph contraction process as well. The elimination process that generates $G_{i+1}$ from $H_i$ naturally generates a spanning tree for $G_{i+1}$. The total stretch of the off-tree edges in $G_{i+1}$ is at most equal to that in $H_i$. This effectively allows us to compute only one low-stretch spanning tree for the first graph in the chain, and keep the same tree for the rest of the chain. This is a significant departure from previous constructions, where a low-stretch spanning tree had to be calculated for \textit{each} $G_i$.

The ability to keep the same low-stretch spanning tree for the whole chain, allows us to prove that Laplacians of \textit{spine-heavy} graphs, i.e. graphs with a spanning tree with average stretch $O(1/\log n)$, can be solved in linear time. This average stretch is a factor of $\tilde{O}(\log^2 n)$ smaller than what is true for general graphs. We reduce the first general graph $G_1$ into a spine-heavy graph $G_2$ by scaling-up the edges of its low-stretch spanning tree by a factor of $\tilde{O}(\log^2 n)$. This results in the construction of a preconditioner chain with a \textit{skewed} set of conditioner numbers. That is, the condition number of the pair $(G_i,H_i)$ is a fixed constant with the exception of $(G_1,H_1)$ for which it is $\tilde{O}(\log^2 n)$. In all previous solvers the condition number for the pair $(G_i,H_i)$ was a uniform function of the size of $G_i$.

An additional significant departure from previous constructions is in the way that the number of edges decreases between subsequent $G_i$'s in the chain. For example, in the \cite{KoutisMP10} chain the number of edges in $G_{i+1}$ is always at least a factor of $\tilde{O}(\log^2 n)$ smaller than the number of edges in $G_i$. In the chain presented in this paper \textit{irregular decreases} are possible; for example a big drop in the number of edges may occur between $G_2$ and $G_3$ and the progress may stagnate for a while after $G_3$, until it starts again.

In order to analyze this new chain we view the graphs $H_i$ as multi-graphs or {\em graphs of samples}.  In the sampling procedure that generates $H_i$, some off-tree edges of $G_i$ can be sampled multiple times, and so $H_i$ is naturally a multi-graph, where the weight of a `traditional' edge $e$ is split among a number of parallel multi-edges with the same endpoints. The progress of the overall sparsification in the chain is then monitored in terms of the number of multi-edges in the $H_i$'s. In other words, when the algorithm appears to be stagnated in terms of the edge count in the $G_i$'s, progress is still happening by `thinning' the off-tree edges. The details are given in Section \ref{sec:solver}.

The final bottleneck to getting an $O(m \log n)$ algorithm for very sparse
systems is the $\tilde{O}(m\log n + n\log^2 n)$
running time of the algorithm for constructing a low-stretch spanning tree
\cite{AbrahamBN08, EEST05}.
We address the problem by noting that it suffices to find a low-stretch spanning tree on a graph with edge weights that are roughly powers of 2.
In this special setting, the shortest path like ball/cone growing routines in
\cite{AbrahamBN08, EEST05} can be sped up in a way similar
to the technique used in \cite{Orlin10}.
We also slightly improve the result of \cite{Orlin10}, which may be of
independent interest.

\section{Background and notation} \label{sec:definitions}

A matrix $A$ is symmetric diagonally dominant if it is symmetric and $A_{ii} \geq \sum_{j \not= i} |A_{ij}|$. It is well understood that any linear system whose matrix is SDD is easily reducible to a system whose matrix is the Laplacian of a weighted graph with positive weights \cite{Gremban-thesis}. The {\em Laplacian} matrix of a graph $G=(V,E,w)$ is the matrix defined as $$L_G(i,j) = -w_{i,j} \textnormal{~and~}  L_G(i,i) = \sum_{j\neq i} w_{i,j}.$$

There is a one-to-one correspondence between graphs and Laplacians which allows us to extend some algebraic operations to graphs. Concretely, if $G$ and $H$ are graphs, we will denote by $G+H$  the graph whose Laplacian is $L_G+L_H$, and by $cG$ the graph whose Laplacian is $cL_G$.

\begin{definition} {\bf [Spectral ordering of graphs]} \\
We define a partial ordering $\preceq$ of graphs by letting
 $$G\preceq H \textnormal{~if and only if~} x^T L_G x \leq x^T L_H x,$$
 for all real vectors $x.~~\bullet$
\end{definition}

If there is a constant $c$ such that $G \preceq cH \preceq \kappa G$, we say that the {\bf condition} of the pair $(G,H)$ is $\kappa$. In our proofs we will find useful to view a graph $G=(V,E,w)$ as a graph with multiple edges.
\begin{definition} {\bf [Graph of samples]} \\
A graph $G=(V,E,w)$ is called a graph of samples, when each edge $e$ of weight $w_e$ is considered as a sum of a set ${\cal L}_e$ of parallel edges, each of weight $w_l=w_e/|{\cal L}_e|$. When needed we will emphasize the fact that a graph is viewed as having parallel edges, by using the notation $G=(V,{\cal L},w).~~\bullet$
\end{definition}

\begin{definition} {\bf [Stretch of edge by tree]} \\
Let $T=(V,E_T,w)$ be a tree. For $e\in E_T$ let $w'_e = 1/{w_e}$. Let $e$ be an edge  not necessarily in $E_T$, of weight $w_e$. If the unique path connecting the endpoints of $e$ in $T$ consists of edges $e_1 \dots e_k$, the stretch of $e$ by $T$ is defined to be $$stretch_T(e) = \frac{\sum_{i=1}^k w'_{e_i}}{w'_e}.~~\bullet$$
\end{definition}

A key to our results is viewing graphs as resistive electrical networks \cite{doyle-2000}. More concretely, if $G=(V,{\cal L},w)$ each $l\in {\cal L}$ corresponds to a resistor of capacity $1/w_l$ connecting the two endpoints of ${\cal L}$. We denote by $R_G(e)$ the {\bf effective resistance} between the endpoints of $e$ in $G$. The effective resistance on trees is easy to calculate; we have $R_T(e) = \sum_{i=1}^k 1/w(e_i) $. Thus $$stretch_T(e)  = w_e R_T(e).$$
We extend the definition to $l \in {\cal L}_e$ in the natural way
$$stretch_T(l)  = w_l R_T(e),$$
and note that $stretch_T(e) = \sum_{l \in {\cal L}_e} stretch_T(l)$.

This definition can also be extended to set of edges. Thus $stretch_T(E)$ denotes the vector of stretch values
of all edges in $E$. We also let  $stretch_T(G)$ denote the vector of stretch for edges in $E_G-E_T$.

\begin{definition} {\bf [Total Off-Tree Stretch] 
} \\
Let $G=(V,E_G,w)$ be a graph, $T = (V,E_T,w)$ be a spanning tree of $G$. We define
$$
    |stretch_{T}(G)| = \sum_{e\in E_G-E_T} stretch_T(e). ~~\bullet
$$
\end{definition}

\section{Incremental Sparsifier} \label{sec:incrementalsparsify}

In their remarkable work \cite{SpielmanS08}, Spielman and Srivastava analyzed a spectral sparsification algorithm based on a simple sampling procedure. The sampling probabilities were proportional to the effective resistances $R_G(e)$ of the edges on the input graph $G$. Our solver in \cite{KoutisMP10} was based on an {\em incremental sparsification} algorithm which used upper bounds on the effective resistances, that are more easily calculated. In this section we give a more careful analysis of the incremental sparsifier algorithm given in \cite{KoutisMP10}.

We start by reviewing the basic \textsc{Sample} procedure. The procedure takes as input a weighted graph $G$ and frequencies $p'_e$ for each edge $e$. These frequencies are normalized to probabilities $p_e$ summing to $1$.  It then picks in $q$ rounds exactly $q$ {\bf samples} which are weighted copies of the edges.  The probability that given edge $e$ is picked in a given round is $p_e$. The weight of the corresponding sample is set so that the expected weight of the edge $e$ after sampling is equal to its actual weight in the input graph. The details are given in the following pseudocode.

\begin{algo}[h]
\qquad

\textsc{Sample}
\vspace{0.05cm}

\underline{Input:} Graph $G=(V,E,w)$, $p':E \rightarrow \R^+$, real $\xi$.

\underline{Output:} Graph $G'=(V,{\cal L}, w')$.
\vspace{0.2cm}

\begin{algorithmic}[1]
\STATE{$t:= \sum_e p'_e$}
\STATE{$q:= C_s  t \log{t} \log(1/\xi)$} {\footnotesize ~~(* $C_S$ is an explicitly known constant  *)}
\STATE{$p_e:= {p'_e}/{t}$}
\STATE{$G':= (V,{\cal L},w')$ with ${\cal L}=\emptyset$}
\FOR{$q$ \mbox{times}}
\STATE{{\small Sample one $e \in E$ with probability of picking $e$ being $p_e$} }
\STATE{{\small Add sample of $e$, $l$ to ${\cal L}_e$ with weight $w'_l = w_e/(p_e q)$}} 
 {\footnotesize ~~~(* Recall that ${\cal L} = \bigcup_{e\in E} {\cal L}_e$ *)}
\ENDFOR
\RETURN{$G'$}
\end{algorithmic}
\end{algo}

 The following Theorem characterizes the quality of $G'$ as a spectral sparsifier for $G$ and it was proved in \cite{KoutisMP10}.

\begin{theorem}\label{thm:sample}

\textbf{(Oversampling)} Let $G=(V,E,w)$ be a graph. Assuming that $p'_e \geq w_eR_G(e)$ for each edge $e\in E$, and $\xi \in \Omega(1/n)$, the graph $G' = \textsc{Sample}(G, p', \xi)$ satisfies
$$
       G \preceq 2G' \preceq 3G
$$
with probability at least $1 - \xi$.
\end{theorem}

\medskip
Suppose we are given a spanning tree $T$ of $G = (V,E,w)$. The incremental sparsification algorithm of \cite{KoutisMP10} was based on two key observations: (a) By Rayleigh's monotonicity law \cite{doyle-2000} we have $R_T(e) \geq R_G(e)$ because $T$ is a subgraph of $G$. Hence the numbers $stretch_T(e)$ satisfy the condition of Theorem \ref{thm:sample} and they can be used in \textsc{Sample}. (b) Scaling up the edges of $T$ in $G$ by a factor of $\kappa$ gives a new graph $G'$ where the stretches of the off-tree are smaller by a factor of $\kappa$ relative to those in $G$. This forces \textsc{Sample} (when applied on $G'$) to sample more often edges from $T$, and return a graph with a smaller number of off-tree edges. In other words, the scale-up factor $\kappa$ allows us to control the number of off-tree edges. Of course this comes at the cost of incurring condition $\kappa$ between $G$ and $G'$.

In this paper we follow the same approach, but also modify \textsc{IncrementalSparsify} so that the output graph is a union of a copy of $T$ and the off-tree samples picked by \textsc{Sample}. To emphasize this, we will denote the edge set of the output graph by $E_T\cup {\cal L}$. The details are given in the following algorithm.

\begin{algo}[h]
\qquad

\textsc{IncrementalSparsify}
\vspace{0.05cm}

\underline{Input:} Graph $G=(V,E,w)$, edge-set  $E_T$ of spanning tree $T$, reals $\kappa>1$, $0<\xi<1$

\underline{Output:} Graph $H=(V,E_T\cup {\cal L})$ or \texttt{FAIL}
\vspace{0.2cm}

\begin{algorithmic}[1]
\STATE{Calculate $stretch_{T}(G)$}
\IF {$|stretch_T(G)| \leq 1$}
\RETURN {$2T$}
\ENDIF
\STATE{$T' := \kappa T$.}
\STATE{$G' := G + (\kappa -1)T$} 
{  ~~~{\footnotesize{(* $G'$ is the graph obtained from $G$  by replacing $T$ by $T'$ *)}}}
\STATE{$\hat{t} := |stretch_{T'}(G')|$}  {~~~~\footnotesize{(* $\hat{t} = |stretch_T(G)|/\kappa$ *)}}
\STATE $t= \hat{t}+n-1$ {~~~~\footnotesize{(* total stretch including tree edges *)}}
\STATE {$\tilde{H}=(V,\tilde{{\cal L}}) := $ \textsc{Sample}($G'$, $stretch_{T'}(E')$, $\xi$)}
\STATE{{\bf if} {$(\sum_{e \not \in E_T} |\tilde{{\cal L}}_e|) \geq 2 (\hat{t}/t)C_s\log t \log(1/\xi)$}} 
{~~~\footnotesize{(* $C_s$ is the constant in \textsc{Sample} *)}}
\STATE {\qquad {\bf return} \texttt{FAIL}}
\STATE{{\bf end}}
\STATE{${\cal L} := \tilde{{\cal L}} - \bigcup_{e\in E_T} {\tilde{{\cal L}}}_e$.}
\STATE{$H := {\cal L} + 3T'$}
\RETURN{$4H$}
\end{algorithmic}
\end{algo}

\begin{theorem} \label{th:incrementalsparsify} Let $G$ be a graph with $n$ vertices and $m$ edges and $T$ be a spanning tree of $G$. Then for  $\xi \in \Omega(1/n)$,
$\textsc{IncrementalSparsify}(G,E_T,\kappa,\xi)$
computes with probability at least $1 - 2\xi$ a graph $H=(V,E_T\cup {\cal L})$ such that
\begin{itemize}
\item
$G  \preceq H \preceq 54 \kappa G$
\item
 $|{\cal L}|\leq 2\hat{t} C_S \log t \log(1/\xi)$
\end{itemize}
where $\hat{t} = stretch_T(G)/\kappa$, $t = \hat{t}+n-1$, and $C_S$ is the constant in \textsc{Sample}. The algorithm can be implemented to run in $\tilde{O} ((n\log{n} + \hat{t} \log^2 n) \log(1/\xi) )$.
\end{theorem}

\Proof
We first suppose that $|stretch_T(G)| \leq 1$ holds.
Thus $G/2 \preceq T \preceq G$, by well known facts~\cite{Boman03support}. Therefore returning $H=2T$ satisfies the claims.
Now assume that the condition is not true. Since in Step 6 the weight of each tree edge is increased by at most a factor of $\kappa$, we have $G \preceq G' \preceq \kappa G$.  \textsc{IncrementalSparsify} sets $p'_e = 1$ if $e \in E_T$ and $stretch_T(e)/\kappa$ otherwise, and invokes \textsc{Sample} to compute a graph $\tilde{H}$ such that with probability at least $1-\xi$, we get
\begin{equation} \label{eq:H}
G\preceq G' \preceq 2\tilde{H} \preceq {3}G' \preceq 3\kappa G.
\end{equation}

We now bound the number $|{\cal L}|$ of off-tree samples drawn by \textsc{Sample}.
For the number $t$ used in \textsc{Sample} we have $t = \hat{t}+n-1$ and $q
= C_s t\log t \log(1/\xi)$ is the number samples drawn by \textsc{Sample}. Let $X_i$ be a random variable which is $1$ if the $i^{th}$ sample picked by \textsc{Sample} is a non-tree edge and
$0$ otherwise. The total number of non-tree samples is the
random variable $X = \sum_{i=1}^q X_i$, and its expected value can be
calculated using the fact $Pr(X_i=1)=\hat{t}/t$:
\begin{eqnarray*}
   E[X] & =  &q \frac{\hat{t}}{t} = \hat{t} \frac{C_s t\log t \log(1/\xi) }{ t} =  C_S \hat{t} \log t \log(1/\xi).
\end{eqnarray*}
Step 12 assures that $H$ does not contain more than $2E[X]$ edges so the claim about the number of off-tree samples is automatically satisfied. A standard form of Chernoff's inequality is:
\begin{eqnarray*}
    Pr[X>(1+\delta)E[X]] & < & exp(-\delta^2E[X]) \\
     Pr[X<(1-\delta)E[X]] & < & exp(-\delta^2E[X]).
\end{eqnarray*}
Letting $\delta =1$, and since $\hat{t}>1,C_S>2$ we get
    $Pr[X>2E[X]]< ( exp({-2}{E[X]}) < 1/n^2$. So, the probability that the algorithm returns a \texttt{FAIL} is at most $1/n^2$. It follows that the probability that an output of \textsc{Sample} satisfies inequality \ref{eq:H} and doesn't get rejected by \textsc{IncrementalSparsify} is at least $1-\xi -1/n^2$.

We now concentrate on the edges of $T$. Any fixed edge $e\in E_T$ is sampled with probability $1/t$ in \textsc{Sample}. Let $X_e$ denote the random variable equal to number of times $e$ is sampled. Since there are $q=C_st\log t \log(1/\xi)$ iterations of sampling, we have $E[X_e]=q/t \geq C_s\log{n}$.
By the Chernoff inequalities above, setting $\delta = 1/2$ we get that
\[
Pr[X_e > (3/2) E[X_e]] \leq exp(- (C_s/4) \log{n})
\]
and
\[
Pr[X_e < (1/2) E[X_e]] \leq exp(- (C_s/4) \log{n}).
\]
By setting $C_s$ to be large enough we get $exp(- (C_s/4) \log{n})<n^{-4}$. So with probability at least $1-1/n^2$ there is no edge $e\in E_T$ such that $X_e>(3/2) E[X_e]$ or $X_e< (1/2) E[X_e]$. Therefore we get that with probability at least $1-1/n^2$ all the edges $e\in E_T$ in $\tilde{H}$ have weights at most three times larger than their weights in $(H/2)$, and
$$
     G\preceq \tilde{H} \preceq  H \preceq 18 \tilde{H}  \preceq 54\kappa G.
$$
Overall, the probability that the output $H$ of \textsc{IncrementalSparsify} satisfies the claim about the condition number is at least  $1-\xi - 2/n^2\geq 1-2/\xi$.

We now consider the time complexity. We first compute the effective resistance of each non-tree edge by the tree.
This can be done using Tarjan's off-line LCA algorithm \cite{Tarjan79}, which takes $O(m)$ time \cite{GabowTarjan83}.
We next call \textsc{Sample}, which draws a number of samples. Since the samples from $E_T$ don't affect the output of \textsc{IncrementalSparsify} we can implement \textsc{Sample} to exploit this; we split the interval $[0,1]$ to two non-overlapping intervals with length corresponding to the probability of picking an edge from $E_T$ and $E-E_T$. We further split the second interval by assigning each edge in $E-E_T$ with a sub-interval of length corresponding to its probability, so that no two intervals overlap. At each sampling iteration we pick a random value in $[0,1]$ and in $O(1)$ time we decide if the value falls in the interval associated with $E-E_T$. If no, we do nothing. If yes, we do a binary search taking $O(\log n)$ time in order to find the sub-interval that contains the value. With the given input \textsc{Sample} draws at most $\tilde{O}(\hat{t} \log n \log (1/\xi))$ samples from $E-E_T$ and for each such sample it does $O(\log n)$ work. It also does $O(n\log n\log (1/\xi) )$ work rejecting the samples from $E_T$. Thus the cost of the call to \textsc{Sample} is $\tilde{O}((n \log n +\hat{t} \log^2 n  )\log (1/\xi))$. \QED

Since the weights of the tree-edges $E_T$ in $H$ are different than those in $G$, we will use $T_H$ to denote the spanning tree of $H$ whose edge-set is $E_T$.
We now show a key property of \textsc{IncrementalSparsify}.
\begin{lemma} {\bf (Uniform Sample Stretch)} \label{lem:sparsified}
  Let $H =(V,E_T\cup {\cal L},w) := \textsc{IncrementalSparsify}(G,E_T,\kappa,\xi)$, and $C_S,t$ as defined in Theorem \ref{th:incrementalsparsify}. For all $l \in {\cal L}$, we have
  $$stretch_{T_H}(l) = \frac{1}{3C_S\log t \log(1/\xi)}.$$
\end{lemma}
\Proof
Let $T' = \kappa T$.  Consider an arbitrary non-tree edge $e$ of $G'$ defined in Step 5 of \textsc{IncrementalSparsify}. The probability of it being sampled is:
\[
p'_e = \frac{1}{t} \cdot w_e \cdot R_{T'}(e)
\]
where $R_{T'}(e)$ is the effective resistance of $e$ in $T'$ and $t= n-1+ s_{T'}(G')=n-1+stretch_T(G)/\kappa$ is the total stretch of all $G'$ edges by $T'$. If $e$ is picked, the corresponding sample $l$ has weight $w_e$ scaled up by a factor of $1/p'_e$,
but then divided by $q$ at the end.
This gives
\begin{eqnarray*}
w_l = \frac{w_e}{p'_e} \cdot \frac{1}{q}
= \frac{w_e}{(w_e R_{T'}(e)) /t}
\cdot \frac{1}{C_S t\log{t}\log(1/\xi)} \\
= \frac{1}{C_S R_{T'}(e) \log{t} \log(1/\xi)}.
\end{eqnarray*}
So the stretch of $l$ with respect to $T'$ is independent from $w_e$ and equal to
\begin{align*}
stretch_{T'}(e) = w_l R_{T'}(e)
=  \frac{1}{C_S \log{t} \log(1/\xi)}.
\end{align*}
Finally note that $T_H = 3T'$. This proves the claim. \QED

\section{Solving using Incremental Sparsifiers} \label{sec:solver}

We follow the framework of the solvers in  \cite{SpielmanTeng08c} and \cite{KoutisMP10} which consist of two phases. The {\em preconditioning phase} builds a chain of graphs ${\cal C} = \{G_1,H_1,G_2,\ldots,H_d\}$ starting with $G_1=G$, along with a corresponding list of positive numbers  ${\cal K} = \{\kappa_1,\ldots,\kappa_{d-1} \}$ where $\kappa_i$ is an upper bound on the condition number of the pair $(G_i,H_i)$. The process for building ${\cal C}$ alternates between calls to a sparsification routine (in our case \textsc{IncrementalSparsify}) which constructs $H_i$ from $G_i$ and a routine \textsc{GreedyElimination} which constructs $G_{i+1}$ from $B_i$, by applying a greedy elimination of degree $1$ and $2$ nodes. The preconditioning phase is independent from the $b$-side of the system $L_Ax=b$. The {\em solve phase} passes $\cal C$, $b$ and a number of iterations $t$ (depending on a desired error $\epsilon$) to the recursive preconditioning algorithm \textsc{R-P-Chebyshev}, described in  \cite{SpielmanTeng08c} or in the appendix of  \cite{KoutisMP10}.

We first give pseudocode for \textsc{GreedyElimination}, which deviates slightly from the standard presentation where the input and output are the two graphs $G$ and $\hat{G}$, to include a spanning tree of the graphs.
\begin{algo}[h]
\qquad

\textsc{GreedyElimination}
\vspace{0.05cm}

\underline{Input:} Graph $G=(V,E,w)$, Spanning tree $T$ of $G$ \\
\underline{Output:} Graph $\hat{G}=(\hat{V},\hat{E},\hat{w})$, Spanning tree $\hat{T}$ of $\hat{G}$
\vspace{0.2cm}

\begin{algorithmic}[1]
\STATE{$\hat{G}:=G$}
\STATE{$E_{\hat{T}}:= E_{T}$}
\REPEAT
\STATE{greedily remove all degree-$1$ nodes from $\hat{G}$}
\IF{$deg_{\hat{G}}(v)=2$ and $(v,u_1),(v,u_2)\in E_{\hat{G}}$}
\STATE{$ w' := \left(1/w(u_1,v)+1/w(u_2,v)\right)^{-1}$}
\STATE{$ w'' := w(u_1,u_2)$ ~~ {\footnotesize (* it may be the case that $w''=0$ *) }}
\STATE{replace the path $(u_1,v,u_2)$ by an edge $e$ of weight $w'$ in $\hat{G}$}
\IF{$(u_1,v)$ or $(v,u_2)$ are not in $\hat{T}$}
\STATE{Let ${\hat{T}} = \{ {\hat{T}} \} - \{(u_1,v), (v,u_2), (u_1,u_2)\}$}
\ELSE
\STATE{Let ${\hat{T}} = \{ {\hat{T}} \cup e \} - \{(u_1,v), (v,u_2), (u_1,u_2)\}$}
\ENDIF
\ENDIF
\UNTIL{there are no nodes of degree $1$ or $2$ in $\hat{G}$ }
\RETURN {$\hat{G}$}
\end{algorithmic}
\end{algo}

Of course we still need to prove that the output $\hat{T}$ is indeed a spanning tree. We prove the claim in the following Lemma that also examines the effect of \textsc{GreedyElimination} to the total stretch of the off-tree edges.
\begin{lemma} \label{th:stretchafterelimination}
Let $(\hat{G},\hat{T}):=\textsc{GreedyElimination}(G,T)$. The output $\hat{T}$ is a spanning tree of $\hat{G}$, and
$$
     |stretch_{\hat{T}}(\hat{G})| \leq |stretch_{T}(G)|.
$$
 \end{lemma}
\Proof
We prove the claim inductively by showing that it holds for all the pairs $(\hat{G}_i,\hat{T}_i)$ throughout the loop, where $(\hat{G}_i,\hat{T}_i)$ denotes the pair $(\hat{G},\hat{T})$ after the $i^{th}$ elimination during the course of the algorithm. The base of the induction is the input pair $(G,T)$ and so the claim holds for it.

When a degree-$1$ node gets eliminated the corresponding edge is necessarily in $E_{\hat{T}}$ by the inductive hypothesis. Its elimination doesn't affect the stretch of any off-tree edge. So, it is clear that if $(\hat{G}_i,\hat{T}_i)$ satisfy the claim then after the elimination of a degree-$1$ node $(\hat{G}_{i+1},\hat{T}_{i+1})$ will also satisfy the claim.

By the inductive hypothesis about $\hat{T}_i$ if $(v,u_1),(v,u_2)$ are eliminated then at least one of the two edges must be in ${\hat{T}_i}$. We first consider the case where one of the two (say $(v,u_2)$) is not in $\hat{T}_i$. Both $u_1$ and $u_2$ must be connected to the rest of $\hat{G}_i$ through edges of $\hat{T}_i$ different than $(u_1,v)$ and $(v,u_2)$. Hence $\hat{T}_{i+1}$ is a spanning tree of $\hat{G}_{i+1}$. Observe that we eliminate at most two non-tree edges from $\hat{G_i}$: $(v,u_2)$ and $(u_1,u_2)$ with corresponding weights $w(v,u_2)$ and $w''$ respectively. Let $\hat{T}[e]$ denote the unique tree-path between the endpoints of $e$ in $\hat{T}$. The contribution of the two eliminated edges to the total stretch is equal to
$$
    s_1 = w(v,u_2) R_{\hat{T}_i}((v,u_2)) + w''R_{\hat{T}_i}((u_1,u_2)).
$$
The two eliminated edges get replaced by the edge $(u_1,u_2)$ with weight $w'+w''$. The contribution of the new edge to the total stretch in $\hat{G}_{i+1}$ is equal to
$$
    s_{2} = w'R_{\hat{T}_{i+1}}((u_1,u_2)) + w''R_{\hat{T}_{i+1}}((u_1,u_2)).
$$
We have $R_{\hat{T}_{i+1}}((u_1,u_2))=R_{\hat{T}_{i}}((u_1,u_2)) <R_{\hat{T}_{i}}((v,u_2))$ since all the edges in the tree-path of $(u_1,u_2)$ are not affected by the elimination. We also have $w(v,u_2)>w'$, hence $s_1>s_{2}$. The claim follows from the fact that no other edges are affected by the elimination, so
\begin{eqnarray*}
   |stretch_{\hat{T}_i}(\hat{G}_i)|-|stretch_{\hat{T}_{i+1}}(\hat{G}_{i+1})|
= s_1-s_{2}>0.
\end{eqnarray*}

We now consider the case where both edges eliminated in Steps 5-13 are in $\hat{T}_i$. It is clear that $\hat{T}_{i+1}$ is a spanning tree of $\hat{G}_{i+1}$. Consider any off-tree edge $e$ not in $\hat{T}_{i+1}$. One of its two endpoints must be different than either $u_1$ or $u_2$, so its endpoints and weight $w_e$ are the same in $\hat{T}_i$. However the elimination of $v$ may affect the stretch of $e$ if $\hat{T}_i[e]$  goes through $v$. Let
\begin{eqnarray*}
    \tau & = & (\sum_{e' \in \hat{T}_i[e]}1/w_{e'}) -(1/w(u_1,v)+1/w(u_2,v))\\ & = & (\sum_{e' \in \hat{T}_{i+1}[e]}1/w_{e'}) - \left(\left(1/w(u_1,v)+1/w(u_2,v)\right)^{-1} + w_e\right)^{-1}.
\end{eqnarray*}
We have
\begin{eqnarray*}
     \frac {stretch_{\hat{T}_i}(e)}{stretch_{\hat{T}_{i+1}}(e)} =
     \frac{w_e \sum_{e' \in \hat{T}_i[e]}1/w_{e'}}{w_e \sum_{e' \in \hat{T}_{i+1}[e]}1/w_{e'} }= \
      \frac{\left(1/w(u_1,v)+1/w(u_2,v)\right)+ \tau}{\left(\left(1/w(u_1,v)+1/w(u_2,v)\right)^{-1} + w_e\right)^{-1} + \tau}  \geq  1.
\end{eqnarray*}
Since individual edge stretches only decrease, the total stretch also decreases and the claim follows.
\QED

A preconditioning chain of graphs must certain properties in order to be useful with \textsc{R-P-Chebyshev}.

\begin{definition} \textbf{[Good Preconditioning Chain]}
\label{dfn:precondchain} \\
 Let ${\cal C} = \{G = G_1,H_1,G_2,\ldots,G_d\}$ be a chain of graphs and ${\cal K} = \{\kappa_1,\kappa_2,\ldots,\kappa_{d-1}\}$ a list of numbers.
We say that $\{{\cal C, K }\}$ is a good preconditioning chain for $G$, if there
exist a list of numbers $\mathcal{U} = \{ \mu_{1}, \mu_{2}, \ldots \mu_{d}\}$
such that:
\begin{enumerate}
   \item $G_i \preceq H_i \preceq \kappa_i G_i$.
   \item $G_{i+1} = \textsc{GreedyElimination}(H_i)$.
   \item $\mu_{i}$ is at least the number of edges in $G_i$.
   \item $\mu_{1}, \mu_{2} \leq m$, where $m$ is the number of edges in $G=G_1$.
   \item $\mu_{i}/\mu_{{i+1}} \geq \lceil c_r \sqrt{\kappa_i} \rceil$ for all $i> 1$ where $c_r$ is an explicitly known constant.
   \item $\kappa_i\geq \kappa_{i+1}$.
   \item $\mu_d$ is a smaller than a fixed constant.
  \end{enumerate}
\end{definition}

Spielman and Teng \cite{SpielmanTeng08c} analyzed the recursive preconditioned Chebyshev iteration \textsc{R-P-Chebyshev} that can be found in the appendix of \cite{KoutisMP10} and showed that the solution of an arbitrary SDD system can be reduced to the computation of a good preconditioning chain. This is captured more concretely by the following Lemma which is adapted from Theorem 5.5 in \cite{SpielmanTeng08c}.
\vspace{0.2cm}
\begin{lemma} \label{th:requirement}
Let $A$ be an SDD matrix with $A=L_G +D$ where $D$ is a diagonal matrix with non-negative elements, and $L_G$ is the Laplacian of a graph $G$.  Given a good preconditioning chain $\{{\cal C,K}\}$ for $G$, a vector ${x}$ such that $||{x}-A^{+}b||_A<\epsilon ||A^{+}b||_A$ can be computed in time
   ${O}(m\sqrt{\kappa_1} + m\sqrt{\kappa_1\kappa_2})  \log(1/\epsilon))$.
\end{lemma}

Before we proceed to the algorithm for building the chain we will need a modified version of a result by  Abraham, Bartal, and Neiman \cite{AbrahamBN08}, which we prove in Section \ref{sec:lowstretch}.
\begin{theorem}
\label{thm:lowstretchtree}
There is an algorithm \textsc{LowStretchTree} that, given a graph $G=(V, E, w)$, outputs a spanning tree $T$ of $G$ such that
\[
\sum_{e \in E} stretch_T(e) \leq O(m\log{n}\log\log^3{n}).
\]
The algorithm runs in $O(m\log{n}+n\log{n}\log\log{n})$ time.
\end{theorem}

Algorithm \textsc{BuildChain} generates the chain of graphs.
\begin{algo}[h]
\qquad

\textsc{BuildChain}
\vspace{0.05cm}

\underline{Input:} Graph $G$, scalar $p$ with $0<p<1$ \\
\underline{Output:} Chain of graphs ${\cal C}= \{G=G_1,H_1,G_2,\ldots,G_d\}$, List of numbers ${\cal K}$.
\vspace{0.2cm}

\begin{algorithmic}[1]
\STATE {\footnotesize (* $c_{stop}$ and $\kappa_c$ are explicitly known constants *)}
 \STATE $G_1 := G$
 \STATE {$T := \textsc{LowStretchTree}(G)$}
 \STATE {$H_1 := G_1 + \tilde{O}(\log^2 n) T$}
 \STATE {$G_2 := H_1$}
 \STATE ${\cal K}:=\emptyset;~{\cal C}:=\emptyset;~i:=2$
 \STATE $\xi := 2\log n$
 \STATE $E_{T_2} := E_T$
 \STATE {\footnotesize (*$n_i$ denotes the number of nodes in $G_i$*)}
 \WHILE{$n_i > c_{stop}$}
 \STATE $ H_i =(V_i,E_{T_i}\cup{\cal L}_i):= \textsc{IncrementalSparsify}(G_i, E_{T_i}, \kappa_c ,p\xi )$
 \STATE  $\{G_{i+1},T_{i+1}\} := \textsc{GreedyElimination}(H_i,T_i)$
 \STATE ${\cal C} = {\cal C}\cup\{G_i,H_i\}$
 \STATE $i: = i+1$
 \ENDWHILE
 \STATE ${\cal K} = \{\tilde{O}(\log^2 n), \kappa_c,\kappa_c,\ldots, \kappa_c\}$
 \RETURN {$\{{\cal C,K}\}$}
\end{algorithmic}
\end{algo}

It remains to show that our algorithm indeed generates a good preconditioning chain.
\begin{lemma} \label{th:incrementalok}
 Given a graph $G$, $\textsc{BuildChain}(G,p)$ produces with probability at least $1-p$, a good preconditioning chain $\{\cal C,K\}$ for $G$, such that  $\kappa_1=\tilde{O}(\log^2 n)$ and for all $i\geq 2$, $\kappa_i=\kappa_c$ for some constant $\kappa_c$. The algorithm runs in time proportional to the running time of $\textsc{LowStretchTree}(G)$.
\end{lemma}
\Proof  Let $l_1$ denote the number of edges in $G$ and $l_i=|{\cal L}_i|$ the number of off-tree samples for $i>1$.  We prove by induction on $i$ that:
\begin{itemize}
\item [(a)] $l_{i+1} \leq 2 l_i/\kappa_c$.
\item [(b)] $stretch_{T_{i+1}}(G_{i+1}) \leq l_i/(C_S \log t_i \log(1/(p\xi)))= \kappa_c \hat{t}_i$, where $C_S,\hat{t}_i$ and $t_i$ are as defined in Theorem \ref{th:incrementalsparsify} for the graph $G_i$.
\end{itemize}

For the base case of $i=1$, by picking a sufficiently large scaling factor $\kappa_1= \tilde{O}(\log^2 n)$ in Step 4, we can satisfy claim (b). By Theorem \ref{th:incrementalsparsify} it follows that $l_2\leq 2l_1/\kappa_c$, hence (a) holds. For the inductive argument, Lemma \ref{lem:sparsified} shows that $stretch_{E_{T_i}}(H_i)$ is at most $l_i/(C_S \log t_i \log(1/(p\xi)))$. Then claim (b) follows from Lemma \ref{th:stretchafterelimination} and claim (a) from Theorem \ref{th:incrementalsparsify}.

We now exhibit the list of numbers $\mathcal{U} = \{ \mu_1, \mu_2 \ldots \mu_{d} \}$
required by Definition \ref{dfn:precondchain}.
A key property of $\textsc{GreedyElimination}$ is that if $G$ is a graph with $n-1+j$ edges, the output $\hat{G}$ of \textsc{GreedyElimination}$(G)$ has at most  $2j-2$ vertices and $3j-3$ edges \cite{SpielmanTeng08c}. Hence the graph $G_{i+1}$ returned by $\textsc{GreedyElimination}(H_i)$ has at most $6l_i/\kappa_c$ edges. Therefore setting $\mu_i=6l_i/\kappa_c$ gives an upper bound on the number of
edges in $G_{i+1}$ and:
$$
    \frac{\mu_i}{\mu_{i+1}} = \frac{6l_i/\kappa_c}{6l_{i+1}/\kappa_c} \geq \frac {3 l_{i+1}}{6l_{i+1}/\kappa_c} \geq \frac{\kappa_c}{2}.
$$
At the same time we have $G_i\preceq  H_i \preceq 54 \kappa_c G_i$. By picking $\kappa_c$ to be large enough we can satisfy all the requirements for the preconditioning chain.

The probability that $H_i$ has the above properties is by construction at least $1-p/(2\log n)$. Since there are at most $2\log n$ levels in the chain, the probability that the requirements hold for all $i$ is then at least
\begin{eqnarray*}
(1-p/(2\log n))^{2\log n} >1-p.
\end{eqnarray*}

Finally note that each call to \textsc{IncrementalSparsify} takes $\tilde{O}(\mu_i\log n \log(1/p))$ time. Since $\mu_i$ decreases geometrically with $i$, the claim about the running time follows. \QED


Combining Lemmas \ref{th:requirement} and \ref{th:incrementalok} proves our main Theorem.

\begin{theorem}
 On input an $n\times n$ symmetric diagonally dominant matrix $A$ with $m$ non-zero entries and a vector $b$,  a vector $x$ satisfying $||x-A^{+}b||_A<\epsilon ||A^{+}b||_A $ can be computed in
 expected time $\tilde{O}(m\log{n}\log(1/\epsilon)).$
\end{theorem}


\section{Speeding Up Low Stretch Spanning Tree Construction} \label{sec:lowstretch}

We improve the running time of the algorithm for finding a low stretch
spanning tree given in \cite{EEST05, AbrahamBN08} by a factor of $\log{n}$,
while retaining the $O(m\log{n}\log\log^3{n})$ bound on total stretch given in
\cite{AbrahamBN08}. Specifically, we claim the following Theorem.
\begin{theorem}
There is an algorithm \textsc{LowStretchTree} that given a graph $G=(V, E, w)$,
outputs a spanning tree $T$ of $G$ in $O(m\log{n}+n\log{n}\log\log{n})$ time
such that
\[
\sum_{e \in E} stretch_T(e) \leq O(m\log{n}\log\log^3{n}).
\]
\end{theorem}

We first show that if the graph only has $k$ distinct edge weights,
Dijkstra's algorithm can be modified to run in $O(m+n\log{k})$ time.
Our approach is identical to the algorithm described in \cite{Orlin10}.
However, we obtain a slight improvement in running time
over the $O(m\log{\frac{nk}{m}})$ bound given in \cite{Orlin10}.

The low stretch spanning tree algorithm in \cite{EEST05,
AbrahamBN08} makes use of Dijkstra's, as well as intermediate
stages of it in the routines \textsc{BallCut} and \textsc{ConeCut}.
We first improve the underlying data structure used by these routines.

\begin{lemma}
\label{lem:fastqueue}
There is a data structure that given a list of non-negative values $L=\{l_1 \dots l_k\}$
(the distinct edge lengths),
maintains a set of keys (distances) starting with $\{0\}$ under the following operations:
\begin{enumerate}
	\item $\textsc{FindMin}()$: returns the element with minimum key.
	\item $\textsc{DeleteMin}()$: delete the element with minimum key.
	\item $\textsc{Insert}(j)$: insert the minimum key plus $l_j$ into the set of keys.
	\item $\textsc{DecreaseKey}(v, j)$: decrease the key of $v$ to the minimum key plus $l_j$.
\end{enumerate}
\textsc{Insert} and \text{DecreaseKey} have $O(1)$ amortized cost
and \textsc{DeleteMin} has $O(\log{k})$ amortized cost.
\end{lemma}

\Proof
We maintain $k$ queues $Q_1 \dots Q_k$ containing the keys with the invariant that
the keys stored in them are in non-decreasing order.
We also maintain a Fibonacci heap as described in \cite{Fredman87} containing
the first element of all non-empty queues.
Since the number of elements in this heap is at most $k$, we can perform
\textsc{Insert} and \textsc{DecreaseKey} in $O(1)$ and \textsc{DeleteMin}
in $O(\log{k})$ amortized time on these elements.
The invariant then allows us to support \textsc{FindMin} in $O(1)$ time.

Since $l_k \geq 0$, the new key introduced by \textsc{Insert} or
\textsc{DecreaseKey} is always at least the minimum key.
Therefore the minimum key is non-decreasing throughout the operations.
So if we only append keys generated by adding $l_j$ to the minimum key
to the end of $Q_j$,
the invariant that the queues are monotonically non-decreasing is maintained.
Specifically, $\textsc{Insert}(j)$ can be performed by appending a new entry
to the tail of $Q_j$.

For $\textsc{DecreaseKey}(v, j)$, suppose $v$ is currently stored in queue
$Q_i$.
We consider two cases:
\begin{enumerate}
\item $v$ has a predecessor in $Q_i$.
Then the key of $v$ is not the key of $Q_i$ in the Fibonacci heap
and we can remove $v$ from $Q_i$ in $O(1)$ time while keeping the invariant.
Then we can insert $v$ with its new key at the end of $Q_j$ using
one \textsc{Insert} operation.

\item $v$ is currently at the head of $Q_i$.
Then simply decreasing the key of $v$ would not violate the invariant of
all keys in the queues being monotonic.
As the new key will be present in the heap containing the first elements
of the queues, a decrease key needs to be performed on the Fibonacci
heap containing those elements.
\end{enumerate}

\textsc{DeleteMin} can be done by doing a delete min in the Fibonacci heap,
and removing the element from the queue containing it.
If the queue is still not empty, it can be reinserted into the Fibonacci
heap with key equaling to that of its new first element.
The amortized cost of this is $O(\log{k}) + O(1) = O(\log{k})$.
\QED

The running times of Dijkstra's algorithm, $\textsc{BallCut}$ and
$\textsc{ConeCut}$ then follows.
\begin{corollary}
\label{lem:fastdijkstra}
Let $G$ be a connected weighted graph and $x_0$ be some vertex.
If there are $k$ distinct values of $d(u, v)$,
Dijkstra's algorithm can compute $d(x_0,u)$ for all vertices $u$
in $O(m+n\log{k})$ time.
\end{corollary}

\Proof
Same as the proof of Dijkstra's algorithm with Fibonacci heap, except
the cost of a \textsc{DeleteMin} is $O(\log{k})$.
\QED

\begin{corollary}
(Corollary 4.3 of \cite{EEST05})
If there are at most $k$ distinct distances in the graph, then
\textsc{BallCut} returns ball $X_0$ such that
\[
cost(\delta(X_0)) \leq O \left( \frac{m}{r_{max} - r_{min}} \right),
\]
in $O(vol(X_0) + |V(X_0)|\log{k})$ time.
\end{corollary}

\begin{corollary}
(Lemma 4.2 of \cite{EEST05})
\label{lem:conecut}
If there are at most $k$ distinct values in the cone distance $\rho$, then

For any two values $0 \leq r_{min} < r_{max}'$,
\textsc{ConeCut} finds a real $r \in [r_{min}, r_{max})$ such that
\begin{eqnarray*}
cost(\delta(B_{\rho}(r, x_0))) \leq \frac{vol(L_r) + \tau}{r_{max} - r_{min}}\cdot  \\
\max \left[ 1, \log_2 \left( \frac{m +\tau}{vol(E(B_{\rho}(r, r_{min})) + \tau }\right) \right],
\end{eqnarray*}
in $O(vol(B_{\rho}(r, x_0)) + |V(B_{\rho}(r, x_0))|\log{k} )$ time,
where $B_{\rho}(r, x_0)$ is the set of all vertices $v$ within distance $r$
from $x_0$ in cone length $\rho$.
\end{corollary}

\Proof
The existence such a $L_r$ follows from Lemma 4.2 of \cite{EEST05}
and the running time follows from the bounds given in Lemma \ref{lem:fastqueue}.
\QED	

We now proceed to show a faster algorithm for constructing low stretch spanning trees by using the data structure from Lemma \ref{lem:fastqueue}. Our presentation is based on the algorithm described in \cite{AbrahamBN08}, which consists of \textsc{HierarchicalStarPartition} at the top level that makes repeated calls to \textsc{StarPartition}.
\textsc{StarPartition} then in turn obtains a desired partition via. calls to \textsc{BallCut} and \textsc{ImpConeDecomp} which uses \textsc{ConeCut}. Due to space limitations we refer to these routines without stating their parameters and guarantees.

\begin{lemma}\label{lem:DecompRuntime}
Given a graph $X$ that has $k$ distinct edge lengths,
The version of \textsc{StarPartition} that uses
\textsc{ImpConeDecomp} as stated in Corollary 6
of \cite{AbrahamBN08} runs in time $O(vol(|X|) + |V(X)|\log{k})$.
\end{lemma}

\Proof
Finding radius and calling \textsc{BallCut} takes
$O(vol(|X|) + |V(X)|\log{k})$ time.
Since the $X_i$s form a partition of the vertices and \textsc{ImpConeDecomp} never reduce the size of a cone, the total cost of all calls to \textsc{ImpConeDecomp} is
\[
\sum_i (vol(X_i) + |V(X_i)|\log{k} )\leq vol(X) + |V(X)| \log{k}.
\]
\QED

We now need to ensure that all calls to \textsc{StarPartition} are made
with a small value of $k$.
This can be done by rounding the edge lengths so that at any iteration of
 \textsc{HierarchicalStarPartition},
the graph has $O(\log{n})$ distinct edge weights.

\begin{algo}[h]
\qquad

\textsc{RoundLengths}
\vspace{0.05cm}

\underline{Input:} Graph $G=(V, E, d)$\\
\underline{Output:} Rounded graph $\tilde{G}=(V, E, \tilde{d})$
\vspace{0.2cm}

\begin{algorithmic}[1]
\STATE{Sort the edge weights of $d$ so that\\
$d(e_1) \leq d(e_2) \leq \dots \leq d(e_m)$.}
\STATE{$i' = 1$}
	\FOR{$i = 1 \dots m$}
		\IF{$d(e_i) > 2d(e_{i'})$}
			\STATE{$i'=i$}
		\ENDIF
		\STATE{$\tilde{d}(e_i) = d(e_{i'})$}
	\ENDFOR
\RETURN{$\tilde{G} = (V, E, \tilde{d})$}
\end{algorithmic}
\end{algo}

The cost of \textsc{RoundLengths} is dominated by the sorting the edges
lengths, which takes $O(m\log{m})$ time.
Before we examine the cost of constructing low stretch spanning tree on $\tilde{G}$,
we show that for any tree produced in the rounded graph $\tilde{G}$,
taking the same set of edges in $G$ gives a tree with similar average stretch.

\begin{claim}
For each edge $e$, $\frac{1}{2}d(e) \leq \tilde{d}(e) \leq d(e)$.
\end{claim}

\begin{lemma}
Let $T$ be any spanning tree of $(V, E)$, and $u, v$ any pair of vertices, we have
\[
\frac{1}{2}d_T(u, v) \leq \tilde{d}_T(u, v) \leq d_T(u, v).
\]
\end{lemma}

\Proof
Summing the bound on a single edge over all edges on the tree path suffices.
\QED

Combining these two gives the following Corollary.
\begin{corollary}
For any pair of vertices $u, v$ such that $uv \in E$,
\[
\frac{1}{2}\frac{\tilde{d}_T(u, v)}{\tilde{d}(u, v)}
\leq \frac{d_T(u, v)}{d(u, v)}
\leq 2\frac{\tilde{d}_T(u, v)}{\tilde{d}(u, v)}.
\]
\end{corollary}

Hence calling $\textsc{HierarchicalStarPartition}(\tilde{G}, x_0, Q)$
and taking the same tree in $G$ gives a low stretch spanning tree for $G$
with $O(m\log{n}\log\log^3{n})$ total stretch.
It remains to bound the running time.
\begin{theorem}
\textsc{HierarchicalStarPartition}$(\tilde{G}, x_0, Q)$ runs in
$O(m \log{m} + n\log{m}\log\log{m})$ time on the rounded graph $\tilde{G}$.
\end{theorem}

\Proof
It was shown in \cite{EEST05} that the lengths of all edges considered at some point
where the farthest point from $x_0$ is $r$ is between $r\cdot n^{-3}$ and $r$.
The rounding algorithm ensures that if $\tilde{d}(e_i) \neq \tilde{d}(e_j)$ for some
$i < j$, we have  $2\tilde{d}(e_i) < \tilde{d}(e_j)$.
Therefore in the range $[r, r \cdot n^{3}]$ (for some value of $r$),
there can only be $O(\log{n})$ different edge lengths in $\tilde{d}$.
Lemma \ref{lem:DecompRuntime} then gives that each call of \textsc{star-partition}
runs in $O(vol(X) + |V(X)|\log\log{n})$ time.
Combining with the fact that each edge appears in at most $O(\log{n})$ layers
of the recursion (Theorem 5.2 of \cite{EEST05}),
we get a total running time of $O(m\log{n}+n\log{n}\log\log{n})$.
\QED

\section{Discussion}

The output of \textsc{IncrementalSparsify} is a graph of samples with a remarkable property as a direct consequence of Lemma \ref{lem:sparsified}; its further incremental sparsification can be performed by a mere {\bf uniform sampling} of its off-tree multi-edges. 

This leads naturally to the definition of a {\bf smooth sequence} of (multi)-graphs on a common set of vertices, with the following properties: (i) it is of logarithmic size, (ii) the first graph is spine-heavy, (iii) every two subsequent graphs have a constant condition number, and (iv) the last graph is a tree. The sequence can be obtained by applying one round of \textsc{IncrementalSparsify} to the spine-heavy graph, and then $O(\log n)$ rounds of uniform sampling.

Smooth sequences of graphs can be useful in an alternative way for building a chain of preconditioners, which separates sparsification from greedy elimination. More concretely, the alternative algorithm first builds a smooth sequence of graphs, starting from the spine-heavy version of the input graph. Then, somewhat roughly speaking, the final chain is obtained by applying a slightly less aggressive version of \textsc{GreedyElimination} to each graph in the sequence; this version eliminates degree-one nodes as usually, but restricts itself to degree-two nodes whose both adjacent edges are in the low-stretch tree. The simplicity of this approach is particularly highlighted in the case of low-diameter unweighted graphs. Solving such graphs has now been essentially  reduced to the computation of a BFS tree followed by a number of rounds of uniform sampling.

We believe that smooth sequences of graphs is a notion of independent interest that may found other applications.

\bibliographystyle{alpha}
\bibliography{Paper.bbl}

\end{document}